\documentclass{article}%
\usepackage{amsmath}
\usepackage{amsfonts}
\usepackage{amssymb}
\usepackage{graphicx}%
\begin{document}

\title{Testing a quintessence model with CMBR peaks location}
\author{D.Di Domenico$^{\dagger}$, C.Rubano$^{\dagger,\bullet}$,
P.Scudellaro$^{\dagger,\bullet}$\\$^{\dagger}$ Dept. of Physical Sciences - Univ. Federico II - Naples\\via Cinthia Napoli Italy\\$^{\bullet}$I.N.F.N - Sec. of Naples\\E-Mail: Rubano@na.infn.it}
\maketitle

\begin{abstract}
We show that a model of quintessence with exponential potential, which allows
to obtain general exact solutions, can generate location of CMBR peaks which
are fully compatible with present observational data.

\end{abstract}

\section{Introduction\hspace{0.05cm}\hspace{0.01cm}}

\hspace{0.01cm}\hspace{0.01cm}In correspondence of some angular scales, the
anisotropy spectrum of the Cosmic Microwave Background Radiation (CMBR)
exhibits some peaks of intensity. Standard cosmology succeeds in explaining
the origin of these peaks and supplies an analytical procedure for the
calculation of their positions. A particularly interesting aspect of this
argument is that such locations are very sensitive to the variation of the
cosmological parameters. Like other kinds of analysis, this places strong
constraints on the cosmological parameters and discriminates among the
different models \cite{1}.

In this paper we want to test a particular model of quintessence. In
order to make this, we first have to calculate, assuming a flat
anisotropy spectrum, the theoretical values of the peaks' positions
and compare these results with the experimental data.

Peaks are formed during recombination age. The perturbative scales
smaller than the value of Hubble's radius at the recombination time
$\eta_{rec}$ re-enter into the horizon before that time and behave
like acoustic waves; in other words, in the primordial plasma they
generate perturbations comparable to those generated by a sound wave
in a fluid \cite{2}.

Small perturbations are therefore generated in matter and radiation, giving
rise to photon clustering.

The radiation pressure acts against this process. The result of these two
competitive processes is the formation of peaks of intensity of radiation
(that is, of peaks of clustering in the distribution of photons) at several
scales, characterized from the relation \cite{3}%

\begin{equation}
k_{n}=\frac{n\pi}{r_{s}(\eta_{rec})}\,, \label{uno}%
\end{equation}
where $\eta$ is the conformal time, $n$ is a positive integer and
$r_{s}$ is the sound horizon at the instant $\eta,$ that is,%

\begin{equation}
r_{s}(\eta)=%
{\displaystyle\int\nolimits_{0}^{\eta}}
c_{s}d\eta^{\prime},
\end{equation}
being $c_{s}$ the sound velocity.

Today we observe the CMBR projected on the sky, so that we are
interested in the angular positions of the peaks. Once we suppose that
the distribution of radiation propagates, from the last scattering up
to now, in such a way that it does not modify the position of the
peaks, these ones are observed at angular scales corresponding to the
three-dimensional scales expressed in Eq. (\ref{uno}). We obtain such
values noticing that to a scale $k_{n}$ corresponds an angular scale
given by%

\begin{equation}
l_{n}\simeq r_{\theta}(\eta_{rec})k_{n}\,,
\end{equation}
where $r_{\theta}$ is the comoving angular distance which in the case
of a spatially flat universe has the simple expression%

\begin{equation}
r_{\theta}=\eta_{0}-\eta_{rec}.
\end{equation}

Consequently, we get%
\begin{equation}
l_{n}=\frac{\eta_{0}-\eta_{rec}}{\overline{c_{s}}\eta_{rec}}n\pi=nl_{A}\,,
\label{rel}%
\end{equation}
where it has been assumed that $c_{s}$ is approximately constant
between $\eta$ and $\eta_{rec}$ and equal to its average value, while
$l_{A}$ is given by%
\begin{equation}
l_{A}\equiv\frac{\eta_{0}-\eta_{rec}}{\overline{c_{s}}\eta_{rec}}\pi\,.
\label{la}%
\end{equation}

\section{Universe with quintessence}

In a model of universe without quintessence or a $\Lambda$-term the
positions of the peaks satisfy relation (\ref{rel}). But the recent
observations give data which cannot be deduced from this relation. On
the other hand, if we consider a model with quintessence, the presence
of this new component modifies Eq. (\ref{rel}). It has been shown that
such modifications can be accounted for introducing a discrete series
of parameters $\varphi_{n}$, in function of which Eq. (\ref{rel}) is
rewritten as \cite{4}%

\begin{equation}
l_{n}=l_{A}(n-\varphi_{n}) \,.
\end{equation}

In \cite{4} the empirical formulas giving the values of the parameters
$\varphi_{n}$ for a generic model of quintessence are deduced. It is%

\begin{equation}
\varphi_{n}=\overline{\varphi}+\Delta\varphi\,,
\end{equation}
with%

\begin{equation}
\overline{\varphi}=a_{1}\left(  r_{rec}\right)  ^{a_{2}}+0.291\overline
{\Omega}_{ls}^{\varphi}\,,
\end{equation}
where
\begin{equation}
a_{1}=0.286+0.262(\Omega_{b}h^{2})\,,
\end{equation}
\begin{equation}
a_{2}=0.1786-6.308(\Omega_{b}h^{2})+174.9(\Omega_{b}h^{2})^{2}-1168(\Omega
_{b}h^{2})^{3}\,,
\end{equation}
\begin{equation}
r_{rec}=\frac{\rho_{r}(z_{rec})}{\rho_{m}(z_{rec})}=0.042\left(  \Omega
_{b}h^{2}\right)  \frac{z_{rec}}{10^{3}}\,,
\end{equation}
\begin{equation}
\overline{\Omega}_{ls}^{\varphi}=\eta_{rec}^{-1}%
{\displaystyle\int\nolimits_{0}^{\eta_{rec}}}
\Omega^{\varphi}(\eta)d\eta\text{ .}%
\end{equation}

In the last one of these relations, it is assumed that $\Omega^{\varphi}%
(\eta)$ does not change too much rapidly before recombination, so that
$\overline{\Omega}_{ls}^{\varphi}$ is an effective averaged value of
quintessential energy.

In \cite{4} Doran and Lilley obtain the formulas that allow to
calculate the theoretical values of the first three peaks.

Therefore, by means of Eq. (\ref{rel}), it is possible to test a given
model of quintessence. In fact, once we fix the values of the
parameters by which the model is specified, we can calculate the
values $l_{n}$ of the angular positions of the peaks and compare them
with the experimental data. In order to perform this calculation, of
course, it is first necessary to find the value of $l_{A}$.

From the definition in Eq. (\ref{la}), and following \cite{4}, we can
use the evolution equations for a universe with quintessence and
derive the expressions for $\eta$ and $\eta_{rec}$ in terms of
$\Omega^{\varphi}$. We can also approximate $\Omega^{\varphi}$ by
means of the constant average $\overline{\Omega}%
_{ls}^{\varphi}$ for the period around last scattering.

In this way, setting $8\pi G=c=1$ and denoting with $\Omega_{0}^{r}$ and
$\Omega_{0}^{\varphi}$ today's radiation and quintessence components, with
$a_{rec}$ the scale factor at last scattering (having we supposed $a_{0}=1$),
and with $\overline{c}_{s}\ $\ the average sound speed before last scattering,
we obtain%

\begin{equation}
3H^{2}(t)\left(  1-\overline{\Omega}_{ls}^{\varphi}\right)  =\rho^{m}%
(t)+\rho^{r}(t)=\rho_{0}^{m}a(t)^{-3}+\rho_{0}^{r}a(t)^{-4}\,. \label{equat}%
\end{equation}

Today, the radiation density is negligible, and we have%

\begin{equation}
3H_{0}^{2}(t)\left(  1-\Omega_{0}^{\varphi}\right)  =\rho_{0}^{m}(t)\,,
\end{equation}
which we insert in Eq. (\ref{equat}) to obtain%

\begin{equation}
\left(  \frac{da}{d\eta}\right)  ^{2}=H_{0}^{2}\left(  1-\overline{\Omega
}_{ls}^{\varphi}\right)  ^{-1}\left[  \left(  1-\Omega_{0}^{\varphi}\right)
a\left(  \eta\right)  +\Omega_{0}^{r}\right]  \,. \label{aeq}%
\end{equation}

Integrating this equation by separation of variables, it is possible to obtain
$\eta(a,H_{0},\overline{\Omega}_{ls}^{\varphi},\Omega_{0}^{\varphi},$
$\Omega_{0}^{r}).$

In particular, we are interested in $\eta_{rec}$, that is,%

\begin{equation}
\eta_{rec}=2H_{0}^{-1}\sqrt{\frac{1-\overline{\Omega}_{ls}^{\varphi}}
{1-\Omega_{0}^{\varphi}}}\left\{  \left(  a_{rec}+\frac{\Omega_{0}^{r}
}{1-\Omega_{0}^{\varphi}}\right)  ^{1/2}-\left(  \frac{\Omega_{0}^{r}
}{1-\Omega_{0}^{\varphi}}\right)  ^{1/2}\right\}  \,.
\end{equation}

If we define $\overline{w}_{0}$ as the $\Omega^{\varphi\text{ }}$ weighted
equation of state of the universe%

\begin{equation}
\overline{w}_{0}=\frac{{\int\nolimits_{0}^{\eta_{0}}}\Omega^{\varphi}%
(\eta)w(\eta)d\eta}{{\int\nolimits_{0}^{\eta_{0}}}\Omega^{\varphi}(\eta)d\eta
}\,,
\end{equation}
we can integrate the cosmological equation for $w\simeq$
cost.$=\overline {w}_{0}$ and rewrite Eq. (\ref{aeq}) as%

\begin{equation}
\left(  \frac{da}{d\eta}\right)  ^{2}=H_{0}^{2}\left[  \left(  1-\Omega
_{0}^{\varphi}-\Omega_{0}^{r}\right)  a\left(  \eta\right)  +\Omega_{0}
^{r}+\Omega_{0}^{\varphi}a(1-3\overline{w}_{0})\right]  \,.
\end{equation}

This equation gives%

\begin{equation}
\eta_{0}=2H_{0}^{-1}\left(  1-\Omega_{0}^{\varphi}\right)  ^{-1/2}F\left(
\Omega_{0}^{\varphi},\overline{w}_{0}\right)  \,.
\end{equation}

Thus, we now have $\eta_{0}$ and $\eta_{rec}$, which we can insert in
Eq. (\ref{la}) for the present time.

The result is \cite{5}%

\begin{equation}
l_{A}=\pi\overline{c}_{s}^{-1}\left[  \frac{F\left(  \Omega_{0}^{\varphi
},\overline{w}_{0}\right)  }{\left(  1-\overline{\Omega}_{ls}^{\varphi
}\right)  ^{-1/2}}\left\{  \left(  a_{rec}+\frac{\Omega_{0}^{r}}{1-\Omega
_{0}^{\varphi}}\right)  ^{1/2}-\left(  \frac{\Omega_{0}^{r}}{1-\Omega
_{0}^{\varphi}}\right)  ^{1/2}\right\}  ^{-1}-1\right]  \,,
\end{equation}
with%

\begin{equation}
F\left(  \Omega_{0}^{\varphi},\overline{w}_{0}\right)  =\frac{1}{2}%
{\displaystyle\int\nolimits_{0}^{1}}
dx\left(  x+\frac{\Omega_{0}^{\varphi}}{1-\Omega_{0}^{\varphi}}
x^{(1-3\overline{w}_{0})}+\frac{\Omega_{0}^{r}(1-x)}{1-\Omega_{0}^{\varphi}
}\right)  ^{-1/2} \,.
\end{equation}

\section{Location of the CMB peaks for an exponential model}

Let us then calculate the locations of the CMB peaks in a particular
model of quintessence. As it will be seen, through the comparison with
the experimental data and the forecasts of the simplest models of
quintessence, the model we use is in good agreement with the
observations.

The model of quintessence here investigated has been studied by Rubano and
Scudellaro \cite{6}, starting from a scalar field with a potential%

\begin{equation}
V(\varphi)=Be^{-\lambda\varphi}\text{ \ ,\ \ \ }\lambda=\sqrt{3/2} \,.
\end{equation}

This choice leads to the following exact solutions for, respectively, the
scale factor, the density and equation of state parameters
\begin{equation}
a=\frac{\tau^{2}(1+\tau^{2})}{\tau_{0}^{2}(1+\tau_{0}^{2})}\,,
\end{equation}%
\begin{equation}
\Omega_{m}=\frac{1+\tau^{2}}{\left(  1+2\tau^{2}\right)  ^{2}}\,,
\end{equation}%
\begin{equation}
w=-\frac{3+2\tau^{2}}{4+4\tau^{2}}\,,
\end{equation}
with $\tau=\omega t,$ $\omega^{2}=\sigma^{2}B^{2}=\left(  3/2\right)  B^{2}$
(this time scale depends on the unknown value of $B$, but this does not affect
the situation presented here, which depends only on $\tau$). The value of
$\tau$ at present epoch is $\tau_{0}$, so that $\Omega_{m0}=\Omega_{m}%
(\tau_{0}).$

The present values of all the relevant parameters, thus, depend only
on the choice of $\tau_{0}$.

Such a model is compatible with the data of the observations of the type-IA
supernovae if we choose \cite{7}

\medskip

$\qquad%
\begin{array}
[c]{l}%
\tau_{0}=\left[  0.82,1.40\right]  \text{ (with the most probable value }%
t_{0}=1.268)\,,\\
\Omega_{m0}\in\left[  0.12,0.30\right]  \text{ (with the most probable value
}\Omega_{m0}=0.15)\,.
\end{array}
$

\medskip

This last result is smaller than $\Omega_{m0}=0.30$, the commonly accepted
one. Indeed, in this model, it is true that $\ \Omega_{m}=0.15$ is the most
probable value, but the distribution statistics of the data gives an
approximately equal probability for the other values of the interval.
Moreover, some experimental evidences exist that justify a consideration of
values of $\Omega_{m0}$ smaller than the usual one (see \cite{9}, for instance).

The model we are testing here has been proved to be compatible also
with the data on the peculiar velocities of galaxies \cite{7b}.

By means of the values given above for $\tau_{0}$ and $\Omega_{m0}$,
it is possible to go back to what is foreseen by this model on the
locations of the peaks. In the calculations it has been placed
$\overline{c}_{s}=0.52$ and $\Omega_{0}^{r}=9.89{\ensuremath{\cdot}}10^{-5}$.

Choosing $\Omega_{b}=0.05$ and $h=0.65$, we obtain, for $l_{1}$, $l_{2}$, and
$l_{3},$ the values in Table I.

The range of the values comes from the range of $\tau_{0}$, at
$1\sigma$ level.

\bigskip%
\[%
\begin{tabular}
[c]{|c|c|c|}\hline & Most Probable Value & Range\\\hline $l_{1}$ &
$232$ & $218-236$\\\hline $l_{2}$ & $572$ & $527-586$\\\hline $l_{3}$
& $884$ & $799-909$\\\hline
\end{tabular}
\]

\ \ \ \ \ \ \ \ \ \ \ \ \ \ \ \ \ \ \ \ \ \ \ \ \ \ \ \ \ \ \ \ \ \ \ \ \ \ \ \ \ \ \ \ Table
I

\bigskip

On the other hand, Table II shows the values of the angular scales at which
CMBR is experimentally observed.

The data in the first column concern the BOOMERANG experiment, while in the
second column the data are obtained from a combined analysis of some
recent observations.

\bigskip%

\[%
\begin{tabular}
[c]{|c|c|c|}\hline
& BOOMERANG Experiment & Combined Data\\\hline
$l_{1}$ & 221$\pm$14 & 222$\pm$14\\\hline
$l_{2}$ & 524$\pm$35 & 539$\pm$21\\\hline
$l_{3}$ & 850$\pm$28 & 851$\pm$31\\\hline
\end{tabular}
\]

\ \ \ \ \ \ \ \ \ \ \ \ \ \ \ \ \ \ \ \ \ \ \ \ \ \ \ \ \ \ \ \ \ \ \ \ \ \ \ \ \ \ \ Table
II

\bigskip

The comparison between the intervals of the values for the single
peaks shows that the exponential model for quintessence reproduces the
observed data enough faithfully. In particular, it turns out to be
more satisfactory when the value of $\tau_{0}$ is a little smaller
than the most probable value, so corresponding to a value of
$\Omega_{m0}$ greater than 0.15, as it is reasonable with respect to
other experimental evidences.

Finally, let us compare our results with the standard theoretical
predictions for models with and without a $\Lambda$-term, which are
showed in Table III.

\bigskip%

\[%
\begin{tabular}
[c]{|c|c|c|}\hline
& Universe with $\overline{w}_{0}=$ $\Omega_{\varphi}=0$ & Universe with
$\overline{w}_{0}=-1$ and $\Omega_{\Lambda}=0.6$\\\hline
$l_{1}$ & $206$ & $220$\\\hline
$l_{2}$ & $502$ & $528$\\\hline
$l_{3}$ & $711$ & $793$\\\hline
\end{tabular}
\]

\ \ \ \ \ \ \ \ \ \ \ \ \ \ \ \ \ \ \ \ \ \ \ \ \ \ \ \ \ \ \ \ \ \ \ \ \ \ \ \ \ \ Table
III

\bigskip

We see that the model without a $\Lambda$-term is out of the
observational ranges (as expected), while the other is fully
compatible with both the data and our results. Therefore, we find no
possibility of distinction.

\section{Conclusions}

Up to now the model presented in \cite{6} has revealed  fully
compatible with observational data. Even if simple exponential
potentials for quintessence do not appear as the preferred ones in the
literature, not giving the most known and used results, they anyway
still seem to be able to describe some phenomena.

As a matter of fact, we have now also found that, concerning CMBR, the
values in Table I are sensibly different from those in Table III.

This gives the hope that more precise measurements of the CMBR
spectrum will be sufficient to discriminate among quintessence models.
(For a discussion on this point see also \cite{10}.)


\begin{thebibliography}{99}                                                                                               %


\bibitem[1]{1}W. Hu and N. Sugiyama, Astrophys. J. 444 (1995) 489

\bibitem[2]{2}V. Mukhanov, H. Feldmann, and R. Branderberger, Phys. Rep. 215
(1992) 203

\bibitem[3]{3}W. Hu, doctoral thesis, astro-ph/9508126

\bibitem[4]{4}M. Doran and M. Lilley, astro-ph/0104486

\bibitem[5]{5}M. Doran et al., Astrophys. J. 559 (2001) 501

\bibitem[6]{6}C. Rubano and P. Scudellaro, Gen. Rel. Grav. 34 (2002) 307

\bibitem[7]{7}M. Pavlov, C. Rubano, M. Sazhin, and P. Scudellaro, Astrophys.
J. 566 (2002) 619

\bibitem[8]{9}N. Bachall, R. Cen, R. Dav\`{e}, J. P. Ostriker, and Q. Yu,
Astrophys. J. 541 (2000) 1

\bibitem[9]{7b}M. Sereno and C. Rubano, astro-ph/0203205, to appeir MNRAS

\bibitem[10]{10}C. Rubano and P. Scudellaro, astro-ph/0203225 to appair Gen.
Rel. Grav.
\end{thebibliography}
\end{document}